\documentclass[12pt]{JHEP3}
\usepackage{amssymb,amsmath,amsfonts,epsfig}
\usepackage{cite}

\newcommand{\beqs}{\begin{equation*}}
\newcommand{\beq}{\begin{equation}}

\newcommand{\eeqs}{\end{equation*}}
\newcommand{\eeq}{\end{equation}}

\newcommand{\beqas}{\begin{eqnarray*}}
\newcommand{\beqa}{\begin{eqnarray}}

\newcommand{\eeqas}{\end{eqnarray*}}
\newcommand{\eeqa}{\end{eqnarray}}

\newcommand{\eq}[2]{\begin{equation} #1 \label{#2} \end{equation}}

\newcommand{\eps}{\varepsilon}
\newcommand{\al}{\alpha}
\newcommand{\be}{\beta}
\newcommand{\ga}{\gamma}
\newcommand{\de}{\delta}
\newcommand{\om}{\omega}

\newcommand{\la}{\lambda}
\newcommand{\si}{\sigma}

\newcommand{\Ga}{\Gamma}
\newcommand{\De}{\Delta}
\newcommand{\Om}{\Omega}

\newcommand{\blist}{\begin{itemize}}

\newcommand{\elist}{\end{itemize}}

\providecommand{\href}[2]{#2}

\DeclareFontFamily{OT1}{rsfs}{}
\DeclareFontShape{OT1}{rsfs}{m}{n}{ <-7> rsfs5 <7-10> rsfs7 <10->rsfs10}{} 
\DeclareMathAlphabet{\mycal}{OT1}{rsfs}{m}{n}

\DeclareMathOperator{\extdm}{d}
\newcommand{\extd}{\extdm \!}

\newcommand{\cO}{{\cal{O}}}

\newcommand{\gtt}{g_{tt}^{(1)}}
\newcommand{\htt}{g_{tt}^{(2)}}

\newcommand{\gtf}{g_{t\varphi}^{(1)}}
\newcommand{\htf}{g_{t\varphi}^{(2)}}

\newcommand{\gty}{g_{ty}^{(1)}}
\newcommand{\hty}{g_{ty}^{(2)}}

\newcommand{\gff}{g_{\varphi \varphi}^{(1)}}
\newcommand{\hff}{g_{\varphi \varphi}^{(2)}}

\newcommand{\gfy}{g_{\varphi y}^{(1)}}

\newcommand{\gyy}{g_{yy}^{(1)}}
\newcommand{\hyy}{g_{yy}^{(2)}}

\newcommand{\ett}{e^{\bar t}_{(1)t}}
\newcommand{\ftt}{e^{\bar t}_{(2)t}}

\newcommand{\etf}{e^{\bar t}_{(1)\varphi}}
\newcommand{\ftf}{e^{\bar t}_{(2)\varphi}}

\title{Lobachevsky holography in conformal Chern--Simons gravity}

\author{Mario Bertin$^a$,  Sabine Ertl$^{b,c}$, Hossein Ghorbani$^{b,d}$, Daniel Grumiller$^b$, Niklas Johansson$^b$ and Dmitri Vassilevich$^{a,e}$\\ 
          $^a$~CMCC, Universidade Federal do ABC,\\
          Rua Santa Ad\'elia, 166,
          Santo Andr\'e, SP Brazil\\ 
          $^b$~Institute for Theoretical Physics, 
          Vienna University of Technology,\\
          Wiedner Hauptstrasse 8--10/136,
          A-1040 Vienna, Austria\\ 
          $^c$~Institute for Theoretical Physics, Karlsruhe Institute of Technology (KIT), \\
          Wolfgang-Gaede-Strasse 1, 76131 Karlsruhe, Germany \\
          $^d$~School of Particles and Accelerators, Institute for Research in Fundamental Sciences (IPM)\\ 
          P.O.~Box 19395-5531, Tehran, Iran\\ 
          $^e$~Department of Theoretical Physics, St Petersburg State University,\\
          St Petersburg, Russia\\ 
           Email: \email{mario.bertin@ufabc.edu.br, sertl@hep.itp.tuwien.ac.at, parsaghorbani@gmail.com, grumil@hep.itp.tuwien.ac.at, niklasj@hep.itp.tuwien.ac.at, dmitry.vasilevich@ufabc.edu.br }}

\abstract{
We propose Lobachevsky boundary conditions that lead to asymptotically $\mathbb{H}^2\times\mathbb{R}$ solutions. As an example we check their consistency in conformal Chern--Simons gravity. The canonical charges are quadratic in the fields, but nonetheless integrable, conserved and finite. The asymptotic symmetry algebra consists of one copy of the Virasoro algebra with central charge $c=24k$, where $k$ is the Chern--Simons level, and an affine $\hat u(1)$. We find also regular non-perturbative states and
show that none of them corresponds to black hole solutions. We attempt to calculate the one-loop partition function, find a remarkable separation between bulk and boundary modes, but conclude that the one-loop partition function is ill-defined due to an infinite degeneracy. We comment on the most likely resolution of this degeneracy.
}


\preprint{TUW--12--25}

\begin{document}

\section{Introduction}

Assuming the holographic principle \cite{'tHooft:1993gx,Susskind:1995vu} is correct then holographic correspondences must also exist between spacetimes that are not asymptotically Anti-deSitter (AdS) and field theories that are not necessarily conformal (CFT). Going beyond the AdS/CFT correspondence \cite{Maldacena:1997re} opens Pandora's box, since there are uncountably many spacetimes that are not asymptotically AdS, and most of them are devoid of interest for physics. When deviating from the canon, it is thus useful to do so as little as possible. With this perspective in mind, a number of interesting holographic correspondences have emerged in that past five years: Schr\"odinger holography \cite{Son:2008ye,Balasubramanian:2008dm,Adams:2008wt,Guica:2010sw}, Lifshitz holography \cite{Kachru:2008yh}, warped AdS holography \cite{Anninos:2008fx,Anninos:2009zi,Compere:2009zj}, deSitter holography \cite{Anninos:2011ui} and flat space holography \cite{Barnich:2006av,Bagchi:2012yk}. In this paper we add to this list of potentially interesting and useful holographic correspondences by considering Lobachevsky holography.

Lobachevsky holography is meant in the sense that spacetime asymptotes to the Lobachevsky plane $\mathbb{H}^2$ times some internal spacetime. 
\eq{
\extd s^2 = \extd \rho^2 + \sinh^2\!\rho\, \extd\varphi^2 + \gamma_{ij} (x^k,\,\rho)\, \extd x^i \extd x^j
}{eq:lob1}
Here $\rho$ is a radial coordinate, $\varphi\sim\varphi+2\pi$ an angular coordinate and $x^i$ are some coordinates of the internal spacetime. In the large $\rho$ limit the internal metric approaches an invertible boundary metric 
$\ga_{ij}^{(0)}$ depending only on the internal coordinates $x^k$.
\eq{
\gamma_{ij} (x^k,\,\rho) = \gamma_{ij}^{(0)} (x^k) + o(1)
}{eq:lob2}

The simplest example --- and the only one considered explicitly in this work --- is when the internal space is a line or an $S^1$, which permits us to use techniques of three-dimensional gravity and two-dimensional field theories.
The main difference to AdS$_2$ holography \cite{Strominger:1998yg,Sen:2011cn} [where the $\sinh^2\!\rho$ in \eqref{eq:lob1} essentially gets replaced by $\cosh^2\!\rho$] is that AdS$_2$ has two disconnected boundary components, while the Lobachevsky plane topologically is a disc.\footnote{The Lobachevsky plane $\mathbb{H}^2$ is sometimes called ``Euclidean AdS$_2$'' and was pictorially represented by M.C.~Escher in his hyperbolic tessellation series ``Circle Limits''. We refrain from using this slightly unfortunate nomenclature since global Euclidean AdS$_2$ has a line-element $\extd s^2=\extd\rho^2+\cosh^2\!\rho\,\extd\varphi^2$ and exhibits two disjoint boundaries at $\rho=\pm\infty$, whereas $\mathbb{H}^2$ has a single boundary at $\rho=\infty$. These are crucial global differences that have important consequences for the holographic description. Of course, locally both spaces are equivalent.}
The simplest theory that has a Lobachevsky solution is conformal Chern--Simons gravity \cite{Afshar:2011yh}. 

This paper is organized as follows. 
In section \ref{se:2} we propose the Lobachevsky boundary conditions.
In section \ref{se:3} we construct the asymptotic symmetry algebra of conformal Chern--Simons gravity with Lobachevsky boundary conditions.
In section \ref{app:np} we discuss non-perturbative states and calculate their canonical boundary charges.
In section \ref{se:4} we perform the one-loop calculation. 
In section \ref{se:5} we conclude.

Our conventions are such that the Levi-Civita symbol satisfies $\epsilon^{t\varphi y} = 1$. Symmetrization is defined as $T_{(\al\be)}=\tfrac12\,(T_{\al\be}+T_{\be\al})$. 

\section{Lobachevsky boundary conditions}\label{se:2}

The line-element \eqref{eq:lob1} can be expanded asymptotically. Using the coordinate $y = 2e^{-\rho}$ instead of $\rho$ we require the metric to fulfill the boundary conditions
\eq{
g_{\mu\nu} = \left(\begin{array}{lll} 
 g_{yy} = 1/y^2 + \cO(1/y) & g_{y\varphi} = \cO(1/y) & g_{y i} =\cO(1) \\
 & g_{\varphi\varphi}= 1/y^2 + \cO(1/y) & g_{\varphi i} = \cO(1) \\
 & & g_{ij}=\ga_{ij}^{(0)} + \cO(y)
\end{array}\right)\,,
}{eq:bcs}
where $\ga_{ij}^{(0)}$ is some invertible matrix with the appropriate signature.
We call the boundary conditions \eqref{eq:bcs} ``Lobachevsky boundary conditions''.

As an example we focus on three spacetime dimensions, where 
\eq{
\ga_{ij}^{(0)}\extd x^i\extd x^j=\pm\extd t^2\,, 
}{eq:lob5}
with the plus (minus) sign referring to Euclidean (Lorentzian) signature.
Our background metric in three dimensions is then given by the global Lobachevsky line-element. 
\eq{
\extd \bar s^2 = \frac{\extd y^2}{y^2} + \frac{\extd\varphi^2}{y^2} (1 - y^2/4)^2 \pm \extd t^2
}{eq:background}
We furthermore denote the sub-leading components as follows
\eq{
g_{tt} = \pm 1 + \gtt y  + \htt y^2 + \ldots \qquad g_{t\varphi} = \gtf + \htf y + \ldots
}{eq:lob3}
and so on. This is thus the form of our Fefferman--Graham expansion, with $g_{\mu\nu}^{(1)}$ ($g_{\mu\nu}^{(2)}$) referring to the first (second) subleading term of the state-dependent contribution to the asymptotic Lobachevsky metric \eqref{eq:bcs}. Further subleading terms denoted by the ellipsis in \eqref{eq:lob3} need not have integer powers in $y$.

In three dimensions the diffeomorphisms $\xi$ that preserve these boundary conditions are given by
\eq{
\xi^t = T(\varphi) + \cO(y) \qquad \xi^\varphi = L(\varphi) + \cO(y^2) \qquad \xi^y = y L'(\varphi) +  \cO(y^2)\,.
}{eq:diffeos}
The functions $T$ and $L$ are only subject to the periodicity condition on $\varphi$, but otherwise free functions of one variable. 

For later purposes we list some geometric identities for the 3-dimensional Lo\-ba\-chev\-sky background \eqref{eq:background}, which can be rewritten as
\eq{
\extd \bar s^2 = g_{ab}^{(2)}\,\extd x^a\extd x^b \pm k_\mu k_\nu \,(\extd x^\mu)^2
}{eq:lob42}
where $g_{ab}^{(2)}\,\extd x^a\extd x^b = \extd\rho^2+\sinh^2\!\rho\,\extd\varphi^2$ is the 2-dimensional Lobachevsky line element and $k^\mu$ a covariantly constant vector field normalized to unity, $k^2=\pm 1$ (in the coordinates above $k=\partial_t$). 
\begin{subequations}
 \label{eq:lob43}
\begin{align}
 \bar R_{\mu\nu} &= \frac12\,\bar g_{\mu\nu}\bar R \pm k_\mu k_\nu \\
 \bar R &= -2 = - \bar R^{\mu\nu}\bar R_{\mu\nu} = \bar R^{\mu\nu}\bar R_\nu^\lambda\bar R_{\lambda\mu} =  R^{(2)} \\
 \bar\nabla_\lambda \bar R_{\mu\nu} &=0 = \bar\nabla_\mu k_\nu = k^\mu \bar R_{\mu\nu} \\
 R_{abcd}^{(2)}&=g_{ad}^{(2)}g_{bc}^{(2)}-g_{ac}^{(2)}g_{bd}^{(2)}
\end{align}
\end{subequations}
The quantities with superscript, like $R^{(2)}$, are defined on the Lobachevsky plane $\mathbb{H}^2$.

The four Killing vectors of the Lo\-ba\-chev\-sky background \eqref{eq:lob42} are given by 
\eq{
 T_0 = i\partial_t \qquad L_0 = i\partial_\varphi \qquad L_{\pm 1} = \pm e^{\pm i\varphi}\,\big(\partial_\rho \pm i \coth\rho\,\partial_\varphi\big)\,.
}{eq:referee1}
They form an $SL(2)\oplus U(1)$ isometry algebra.
\eq{
[L_0,\,L_{\pm 1}] = \mp L_{\pm 1}\qquad [L_1,\,L_{-1}] = 2L_0 \qquad [T_0,\,L_n] = 0
}{eq:referee2}

\section{Asymptotic symmetry algebra}\label{se:3}

The boundary conditions presented in the previous section do not depend on any particular theory; however, not every gravitational theory consistently supports a Lobachevsky background. In this section we choose Lorentzian signature and focus on 
conformal Chern--Simons gravity.

In subsection \ref{se:2.1} we present the action and explain why one should expect Lobachevsky holography to work for this theory.
In subsection \ref{se:2.2} we derive the canonical charges and show that they are non-trivial, integrable, finite and conserved.
In subsection \ref{se:2.3} we study the action of pure gauge transformations and consider descendants of the vacuum.
In subsection \ref{se:2.5} we provide the asymptotic symmetry algebra, including its central charges.

\subsection{Conformal Chern--Simons gravity}\label{se:2.1}

Lobachevsky space can be obtained as a $\nu\to 0$ limit from warped AdS, where $\nu$ is the warping parameter (see e.g.~\cite{Anninos:2008fx}).
In topologically massive gravity \cite{Deser:1982vy} 
the warping parameter $\nu$ scales with the Chern--Simons coupling $\mu$ in this limit. This suggests that topologically massive gravity in the scaling limit $\mu\to 0$ should support Lobachevsky solutions. This limit leads to conformal Chern--Simons gravity, which indeed has such solutions \cite{Afshar:2011yh}. 
This theory is topological, in the sense that it has zero local physical degrees of freedom, and appears to be the simplest purely gravitational theory permitting the study of Lobachevsky holography.

The conformal Chern--Simons gravity action
\eq{
S_{\textrm{\tiny CSG}}[g]=-\frac{k}{4\pi}\,\int\!\extd^3x\, \epsilon^{\la\mu\nu}\,\Ga^\si{}_{\la\rho}\,\Big(\partial_\mu\Ga^\rho{}_{\nu\si}+\tfrac23\,\Ga^\rho{}_{\mu\tau}\Ga^\tau{}_{\nu\si}\Big)
}{eq:gCS10}
contains one coupling constant, the Chern--Simons level $k$. 
Besides diffeomorphism invariance in the bulk the theory described by the action \eqref{eq:gCS10} also enjoys invariance under local Weyl rescalings of the metric, \eq{
g\to e^{2\Om} g\,, 
}{eq:lob15}
with some Weyl factor $\Om$ that asymptotically vanishes linearly, $\Om = \cO(y)$, due to our boundary conditions \eqref{eq:bcs}. The equations of motion descending from the action \eqref{eq:gCS10} are solved if and only if the Cotton tensor vanishes,
$C_{\mu\nu}=0$.
Thus, all conformally flat spacetimes are classical solutions of conformal Chern--Simons gravity and vice versa.

\subsection{Canonical charges}\label{se:2.2}

To compute the charges corresponding to the gauge transformations found in the previous sections we use the
expressions for their variation derived in \cite{Afshar:2011yh}. These are obtained in the first order formulation and 
expressed in terms of the Dreibein $e^i{}_\mu$ related to the metric as $g_{\mu\nu} = e^i{}_\mu e^j{}_{\nu} \eta_{ij}$. Therefore, it is useful to provide a Fefferman--Graham expansion in terms of this quantity. The most general expansion resulting in \eqref{eq:bcs} is
\eq{
e^i{}_\mu = \left(\begin{array}{lll} 
 e^{\bar y}{}_y = 1/y + \cO(1) & e^{\bar y}{}_\varphi=\cO(1) & e^{\bar y}{}_t=\cO(y) \\
 e^{\bar\varphi}{}_y = \cO(1) & e^{\bar\varphi}{}_\varphi=1/y + \cO(1) & e^{\bar\varphi}{}_t=\cO(y) \\
 e^{\bar t}{}_y=\cO(1) & e^{\bar t}{}_\varphi=\cO(1) & e^{\bar t}{}_t=-1 + \cO(y) 
\end{array}\right)\, .
}{eq:bcs_e}
Just as for the metric, we assume a Fefferman--Graham expansion of the Dreibein components:
\eq{
e^{\bar t}{}_t = -1 +  y\; \ett + y^2\; \ftt + \ldots \qquad 
e^{\bar t}{}_\varphi = \etf + y\; \ftf + \ldots
}{eq:exp_e}
Note that using Lorentz invariance, we could restrict the form of \eqref{eq:bcs_e} further, but we
prefer to keep the Lorentz gauge unfixed.

Now, the expressions for the variations of the diffeomorphism charges\footnote{In conformal Chern--Simons gravity there are also conserved Weyl charges. In the present case we impose boundary conditions that require the asymptotic Weyl factor to vanish at least linearly in $y$, which leads to vanishing Weyl charges.} are \cite{Afshar:2011yh}
\begin{equation}\label{eq:diffcharge}
\de Q[\xi^\mu] = \frac{k}{2\pi}\,\int\limits_0^{2\pi} \extd \varphi \, \Big[ \xi^\mu \big(
e^i{}_\mu\, \de \la_{i\varphi}  + \la^i{}_\mu\, \de e_{i\varphi}  + 2 \om^i{}_\mu\, \de \om_{i\varphi} \big) +  2\theta^i\, \de\om_{i\varphi} \Big]\, .
\end{equation}
In these expressions $\om^i{}_\mu$ is the spin connection and $\la^i{}_\mu$ is a Lagrange multiplier.
They are given by the torsion constraint $\extd e^i+\eps^i{}_{jk}\,\om^j e^k=0$ and equations of motion $\la_{mn} = -2 \big(R_{mn}-\frac{1}{4}\eta_{mn}R\big)$.
The Ricci tensor and scalar are given by standard identities. 
In the present case it turns out that the contribution from the Lorentz parameters $\theta^i$ to the charges vanishes, so that the last term in \eqref{eq:diffcharge} can be dropped.

After a straightforward but lengthy calculation we obtain the result for the diffeomorphism charges,
\eq{
Q[\xi^\mu] = \frac{k}{2\pi}\,\int\limits_0^{2\pi} \extd \varphi \, \left[T(\varphi) \,\big(\partial_t \gfy - 2 \gtf \big) + L(\varphi) \,f(g^{(1)},\,g^{(2)})\right]
}{eq:lob6}
with
\eq{
\begin{split}
f(g^{(1)},\,g^{(2)}) =&\; \partial_t\partial_\varphi \gtf - 2\partial_\varphi \gfy \\ 
& + 3 \htt+ 2 \hff + \hyy - 3 \partial_t \hty -\frac{1}{2}\partial_t^2 \hff +\frac{1}{2} \partial_t^2 \hyy  \\ 
& + \frac{5}{4} (\gtt)^2 - \frac{1}{2} (\gff)^2 - (\gyy)^2 + 3 (\gtf)^2 + (\gty)^2 - (\gfy)^2 \\ 
& + \frac{3}{4} \gtt \gff - \frac{5}{4} \gtt \gyy - \frac{1}{4} \gff \gyy \\ 
& - \big(2 \gtt  + \frac{3}{2} \gff - \frac{5}{2} \gyy \big)\,\partial_t \gty -3 \gtf \partial_t \gfy +3 \gfy \partial_t \gtf  \\ 
& - \gty \partial_t \gtt + \frac{3}{2} \gty \partial_t \gff + \frac{1}{2} \gty\partial_t \gyy \\ 
& + \frac{1}{2}(\partial_t \gff)^2 -\frac{1}{4}(\partial_t \gyy)^2 +\frac{1}{2}(\partial_t \gfy)^2 \\ 
& - \frac{1}{4} (\partial_t \gtt)(\partial_t\gff)+\frac{1}{4}(\partial_t \gtt)(\partial_t \gyy) - \frac{1}{4} (\partial_t \gff) (\partial_t \gyy) \\ 
& + \frac{1}{2} \big(\gtt + \gff - \gyy\big)\,\partial_t^2 \gyy - \frac{1}{2} \gtt \partial_t^2 \gff - \gfy \partial_t^2 \gfy   \,. 
\end{split}
}{eq:lob7}

We summarize some properties of the charges \eqref{eq:lob6}, \eqref{eq:lob7}:
\begin{itemize}
 \item The charges depend not only (quadratically) on the linearized fluctuations $g^{(1)}$, but also (linearly) on the next order $g^{(2)}$.
 \item The charges are integrable, despite of the appearance of bi-linear terms.
 \item The charges are manifestly finite.
 \item The charges are conserved in time as a consequence of the asymptotic equations of motion (see appendix \ref{app:eom}).
\end{itemize}
The properties above, in particular the first two items, are also true for Comp{\`e}re--Detournay boundary conditions \cite{Compere:2009zj} for asymptotically warped AdS spacetimes \cite{Anninos:2008fx} in topologically massive gravity \cite{Deser:1982vy}. 

Evaluating the diffeomorphism charges for the background \eqref{eq:background}, we realize that all expansion coefficients are zero except $\hff = -1/2$. 
Therefore the first term in the charges \eqref{eq:lob6} vanishes, while the second one, with $f(0,\,\bar g^{(2)}) = -1$, leads to
\eq{
\bar Q[\xi^\mu] = -\frac{k}{2\pi}\,\int\limits_0^{2\pi} \extd \varphi \, L(\varphi)\, .
}{eq:lob9}
This means that the $L$-charge zero-mode is nonzero for the vacuum whereas all other charges are zero.
This zero-mode corresponds to the angular momentum $J = Q[\partial_\varphi]$, and thus we have
\eq{
\bar J = \bar Q[\partial_\varphi] = -k\,. 
}{eq:lob10}
Thus, the background \eqref{eq:background} has a Casimir angular momentum that equals minus the Chern--Simons level.
Other exact backgrounds and their canonical charges are discussed in section \ref{app:np} below.

\subsection{Action of gauge transformations and vacuum descendants}\label{se:2.3}

As a consistency-check we demonstrate that the canonical charges are invariant under trivial gauge transformations.
It turns out that these have a fairly complicated action on the components of $g^{(1)}_{ij}$ and $g^{(2)}_{ij}$. We present below the action on $g^{(1)}_{ij}$ but give just two examples of $g^{(2)}_{ij}$.

We consider a diffeomorphism generated by a vector field $\xi$ with components
\eq{\begin{split}
\xi^t &= T_1 y + T_2 y^2 + T_3 y^3  \\ 
\xi^\varphi &= L_1 y^2 + L_2 y^3 + L_3 y^4 \\ 
\xi^y &= H_1 y^2 + H_2 y^3 + H_3 y^3 \,
\end{split}
}{eq:gaugetrafo}
and a Weyl rescaling \eqref{eq:lob15} with Weyl factor
\eq{
\Omega = \omega_1 y + \om_2 y^2 + \om_3 y^3\, .
}{eq:lob11}
All expansion coefficients $T_i$, $L_i$, $H_i$ and $\om_i$ are functions of $t$ and $\varphi$.
Under this gauge transformation the metric components transform as 
\begin{subequations}
 \label{eq:trans}
\begin{align}
\de \gtt &= -2 \partial_t T_1 - 2 \omega_1  &\de \gtf &= \partial_t L_1  \\
\de \gty &= -T_1 + \partial_t H_1  &\de \gff &= -2 H_1 + 2 \omega_1   \\
\de \gfy &=2 L_1 &\de \gyy &=2 H_1 + 2 \omega_1\, . 
\end{align}
\end{subequations}
Two representative example of the subleading components are
\eq{
\begin{split}
\de \hff &= 2 \partial_\varphi L_1 +2 \omega_2 - 2 H_2  + T_1 \partial_t \gff  + 2 \omega_1 \gff - H_1 \gff \\
\de \hyy &=  4 H_2 + 2 \omega_2 + T_1 \big(\partial_t \gyy + 2 \gty\big) + 3  H_1 \gyy + 2\omega_1 \gyy + 4   L_1 \gfy \, .
\end{split}
}{eq:lob12}
Neither $T_3$, $L_3$, $H_3$ nor $\om_3$ contributes to any component of $\de g^{(2)}_{ij}$.

From \eqref{eq:trans} it is clear that the charges corresponding to $T(\varphi)$ are gauge invariant. Checking the transformation properties of $f$
is slightly lengthy, but straightforward. It turns out that $f$ is not invariant by itself, but transforms by a 
combination of the equations of motion. The quantity
\eq{
f_{\rm inv} = f - \frac{1}{2} (\gff - \gyy) \, {\rm eom}_{t\varphi} 
+\big(2 \gty + \frac{1}{2} \partial_t (\gff - \gyy ) \big) \partial_t \, (\textrm{eom})_{t\varphi} 
 }{eq:lob13}
is off-shell gauge invariant, with $\textrm{eom}_{t\varphi}$ as defined in \eqref{eq:app1}.

It is worthwhile noting that the terms in $f$ linear in $g^{(1)}_{ij}$ are not invariant on their own.
Also the full linear term is not invariant. Thus, the quadratic contributions to $f$ are essential for consistency.

It is possible to exploit small gauge transformations generated by \eqref{eq:gaugetrafo} to set to zero most of the metric components in $g^{(1)}$ and $g^{(2)}$ and thereby considerably simplify the expression for the canonical charges. Namely, by choosing suitably the functions $T_{1,2}$, $L_{1,2}$, $H_{1,2}$ and $\omega_{1,2}$ we can always impose the gauge-fixing conditions
\eq{
g^{(1,2)}_{ty} = g^{(1,2)}_{\varphi y} = g^{(1,2)}_{\varphi\varphi} = g^{(1,2)}_{yy} = 0\,.
}{eq:simplegauge}
The on-shell condition \eqref{eq:app1} then additionally sets to zero $g_{tt}^{(1)}$, while \eqref{eq:app2} [\eqref{eq:app3}] requires $g_{t\varphi}^{(1)}$ [$g_{tt}^{(2)}$] to depend on $\varphi$ only. In this gauge and going on-shell the charges \eqref{eq:lob6}, \eqref{eq:lob7} simplify to
\eq{
Q[\xi^\mu] = \frac{k}{2\pi}\,\int\limits_0^{2\pi} \extd \varphi \, \left[-2T(\varphi) \,\gtf(\varphi) + 3 L(\varphi) \,\big(g_{tt}^{(2)}(\varphi) + (g_{t\varphi}^{(1)}(\varphi))^2\big)\right]\,.
}{eq:simplecharges}
They are manifestly conserved, $\partial_t Q=0$. A slightly more complicated but otherwise similar gauge choice fixes $g^{(1,2)}_{tt} = g^{(1,2)}_{ty} = g^{(1,2)}_{\varphi y} = g^{(1,2)}_{yy} = 0$. If one demands $\tau$-independence of $\gff$ the same statements as above hold, with $\htt$ replaced by $\tfrac23\,\hff$.

When acting on the vacuum with the asymptotic symmetry group, non-trivial linearized states are generated. 
In fact, acting with the diffeomorphism 
\eq{
\xi = T(\varphi)\partial_t + L(\varphi) \partial_\varphi + y L'(\varphi)\partial_y \,
}{eq:ASG}
on the metric \eqref{eq:background} produces a state $g_{\mu\nu} = \bar g_{\mu\nu} + h_{\mu\nu}$ with
\eq{
h_{\mu\nu} = \left(\begin{array}{lll}
  h_{yy}= 0 & h_{y\varphi}=L''(\varphi)/y & h_{yt}= \cO(y^2)\\
   & h_{\varphi\varphi}=-L^\prime(\varphi)+\cO(y^2) & h_{\varphi t}=-T^\prime(\varphi)+\cO(y^3)\\
  & & h_{tt} = \cO(y^3)\\ 
\end{array}\right)\,.
}{eq:lob14}
The corresponding $T$-charges are nonzero. We refrain from computing the quantity $f$ for the descendants since this
quantity is unlikely to make sense for linearized solutions. The fact that $\gfy = L''(\varphi)$ also completes the argument that the linear term in $f$ is not
gauge invariant on its own, while the full expression for $f$ \eqref{eq:lob7} is on-shell gauge invariant [and \eqref{eq:lob13} is even off-shell gauge invariant].

\subsection{Central extensions}\label{se:2.5}

Let us now present the full asymptotic symmetry algebra including central extensions.
The computations follow the standard procedure, which for conformal Chern--Simons gravity is performed in detail in \cite{Afshar:2011yh}.
We define our algebra generators as
\eq{
T_n = \tilde G[T(\varphi) = e^{in\varphi};\, L(\varphi)=0] \qquad L_n = \tilde G[T(\varphi) = 0;\, L(\varphi)=e^{in\varphi}]\,,
}{eq:lob16}
where $\tilde G$ is the canonical generator of Poincar\'e transformations, including the boundary piece from the canonical charge \eqref{eq:lob6}.
Replacing in the end Dirac brackets by commutators, $i\{,\}\to[,]$, and suitably shifting the zero mode generator $L_0$ we obtain finally the asymptotic symmetry algebra.
\begin{subequations}
 \label{eq:lob17}
\begin{align}
 [T_m,\, T_n] &= 2 k \,m \,\delta_{m+n,\,0}\\
 [T_m,\, L_n] &= m T_{m+n}\\
 [L_m,\, L_n] &= (m-n)L_{m+n} + 2 k \,m(m^2-1) \,\delta_{m+n,\,0}
\end{align}
\end{subequations}
We see that the central charge of the Virasoro algebra is 
\eq{
c = 24k\,.
}{eq:lob18}
The result \eqref{eq:lob18} for the central charge is consistent with the $\mu\to 0$ (and $8G\mu\to 1/k$) limit of the right central charge appearing in warped AdS holography, $c_R = (15 \mu^2\ell^2+81)/[G\mu(\mu^2\ell^2+27)]$ \cite{Anninos:2008fx}.
As expected, the centerless subalgebra of the asymptotic symmetry algebra \eqref{eq:lob17} generated by $T_0, L_0, L_{\pm 1}$ coincides with the isometry algebra \eqref{eq:referee2} of the Lobachevsky background.

In conclusion, the asymptotic symmetry algebra \eqref{eq:lob17} consists of an affine $\hat u(1)$ algebra generated by $T_n$ and a Virasoro algebra generated by $L_n$. The central charge is positive provided the level $k$ is positive, with the overall sign choice of the action as in \eqref{eq:gCS10}. Our results suggest that the dual field theory is a warped CFT \cite{Detournay:2012pc}.

\section{Non-perturbative states and their charges}\label{app:np}

In this section we discuss non-perturbative states --- exact metric backgrounds that
solve the equations of motion and are smooth and regular, at least outside possible
black hole horizons. We restrict ourselves exclusively to stationary and axi-symmetric solutions.
By comparison, in AdS$_3$ holography such states are the BTZ
black holes \cite{Banados:1992}. In the present case, however, we shall demonstrate that there are no
regular and smooth black hole solutions. Nevertheless, we are able to identify three non-trivial
non-perturbative states and calculate their canonical boundary charges.

The line-element 
\eq{
\extd s^2 = -\extd t^2 + 2A \extd t\extd\varphi + \big(r^2-Br\big)\extd\varphi^2 + \frac{\extd r^2}{r^2-Br+A^2+C} 
}{eq:cbh1}
for any values of $A,\, B$ (and with $C=0$) is a solution of $C_{\mu\nu}=0$ that asymptotes to $\mathbb{H}^2\times\mathbb{R}$ in the large $r$ limit (for non-vanishing $C$ the relation $B^2=4A^2$ must hold). Note that all curvature invariants are constant and coincide with the ones of Lobachevsky spacetime \eqref{eq:lob43}. Moreover, the line-element \eqref{eq:cbh1} after the coordinate redefinition $r=1/y+B/2+(B^2-4A^2-4C)\,y/16$ is manifestly compatible with our boundary conditions \eqref{eq:bcs}. 
In the notations of sections \ref{se:2} and \ref{se:3} we obtain the non-vanishing expansion coefficients
\eq{
 g_{t\varphi}^{(1)} = A \qquad
 g_{\varphi\varphi}^{(2)} =-\frac{4A^2+B^2+4C}{8} \,.
}{eq:cbh6}
Thus, the zero mode charges are given by
\eq{
M = Q[\partial_t] = -2kA\qquad J=Q[\partial_\varphi] = k \,\big(2A^2-\frac{B^2}{4}-C\big)\,.
}{eq:cbh7}

The non-perturbative states generated by the line-element \eqref{eq:cbh1} could be relevant states in the dual field theory, unless they have to be ruled out for physical reasons. We show now that indeed nearly all of these states are ruled out because they correspond to geometries with naked closed time-like curves (CTCs). 


CTCs emerge unless a) there is a double zero in the $\extd\varphi^2$ term and no coincident pole in the $\extd r^2$ term (solution with a center), b) there is a double zero in the $\extd\varphi^2$ term and a coincident double pole in the $\extd r^2$ term (Poincar\'e horizon), c) there is a single zero in the $\extd\varphi^2$ term and a coincident single pole in the $\extd r^2$ term (solution with center), d) there is no zero in the $\extd\varphi^2$ term (solution with second asymptotic region). We disregard possibility d) since it violates our assumption about cylindrical topology. [Solitonic solutions of this type can be brought into the form $\extd s^2 = -\extd t^2 + 2A \extd t\extd\varphi + \big(r^2+B^2\big)\extd\varphi^2 + \extd r^2/(r^2+A^2+B^2)$ with $M=-2kA$ and $J=k\,(2A^2+B^2)$.] 

Let us focus first on the case $C=0$.
Possibility a) requires $B=0$ and generically leads to a solution with conical defect. The only exception arises if additionally $A=\pm 1$ holds.
Possibility b) requires $A=B=0$. 
Possibility c) requires $A=0$ and generically leads to a solution with conical defect. The only exception arises if additionally $B=\pm 2$ holds. 

A similar analysis can be performed for $C\neq 0$, which implies $B=\pm 2A$ from the equations of motion $C_{\mu\nu}=0$. Possibility a) requires $A=B=0$ and the absence of conical defects sets $C = 1$. Possibility b) does not exist for $C\neq 0$. Possibility c) requires negative $C$ and $B=\pm 2\sqrt{-C}$. The absence of conical defects sets $C=-1$. 

In terms of the canonical charges all the regular states with $C\neq 0$ coincide with some states with $C=0$. Thus, to classify all allowed states in terms of mass $M$ and angular momentum $J$ it is sufficient to consider the cases a), b) and c) for vanishing $C$. Moreover, it is sufficient to require $B$ to be non-negative, since it appears only quadratically in the charges \eqref{eq:cbh7}.
According to the analysis above, there are four different regular non-perturbative states for $C=0$, $B\geq 0$:
\begin{subequations}
\label{eq:np}
\begin{align}
 & \textrm{Global\;Lobachevsky:} && (A=0, B=2) & M&=0 & J &= -k \\
 & \textrm{Poincar\'e\;Lobachevsky:} && (A=B=0) & M &= 0 & J &= 0 \\
 & \textrm{Rotating\;Lobachevsky:} && (A=\pm 1, B=0) & M &=\pm 2k & J &= 2k 
\end{align}
\end{subequations}
Global Lobachevsky is our vacuum state. The other three are non-vacuum states.


Up to trivial gauge transformations, we have not found any additional solutions besides \eqref{eq:cbh1}. It seems plausible
to us that there are no further solutions with cylindrical topology, except for singular ones or solutions that
are gauge-equivalent to the ones presented in this section. Therefore, we conjecture
that the four states listed in \eqref{eq:np} comprise all (regular, stationary and axi-symmetric) non-perturbative states.

\section{One-loop calculations}\label{se:4}

In this section we analyze the one-loop partition function for conformal Chern--Simons gravity \eqref{eq:gCS10}  with Lobachevsky boundary conditions \eqref{eq:bcs}, \eqref{eq:lob5}, with the idea to compare with some corresponding field theoretical partition function, similar to the Einstein gravity precursor with Brown--Henneaux boundary conditions \cite{Maloney:2007ud,Giombi:2008vd,David:2009xg}.  

Here we use Euclidean signature, work on $\mathbb{H}^2\times S^1$, and employ the same strategy as applied earlier for the one loop calculations in three-dimensional gravity \cite{Gaberdiel:2010xv,Bertin:2011jk} (and much earlier in four dimensions \cite{Vassilevich:1992rk}).
We subdivide all metric fluctuations into fluctuations $h_{\rm gf}$ satisfying some gauge fixing condition and
pure gauge fluctuations $h_{\rm gauge}$ parametrized by the gauge group parameters $\zeta$, 
$h=h_{\rm gf}+h_{\rm gauge}(\zeta)$. In this formalism, the ghost factor is given by the Jacobian
in the change of the variables: $\mathcal{D}h=Z_{\rm gh}\mathcal{D}h_{\rm gf}\mathcal{D}\zeta$.
To obtain the one-loop partition function, one has to truncate the classical action to the second
order in fluctuations, $S_2$, and evaluate the path integral  
\eq{
Z=\int \mathcal{D} h\, e^{-S_2(h)} = \int Z_{\rm gh}\mathcal{D}h_{\rm gf}\mathcal{D}\zeta\, e^{-S_2(h)} =Z_{\rm gh} \int \mathcal{D}h_{\rm gf}\, e^{-S_2(h_{\rm gf})}\,,
}{eq:pint}
where we used that $S_2$ does not depend on the gauge parameters $\zeta$, so that the corresponding integration
may be performed giving an irrelevant (infinite) constant equal to the volume of the gauge group.

The remainder of this section is organized as follows.
In subsection \ref{se:3.1} we calculate the second variation of the classical action around the (Euclidean) Lobachevsky background and derive the one-loop determinant of the gauge-fixed fluctuations $h_{\rm gf}$.
In subsection \ref{se:3.2} we evaluate the ghost determinant.
In subsection \ref{se:3.3} we consider boundary conditions for the physical and ghost modes, relying on the analysis of section \ref{se:2} and \ref{se:3}.
In subsection \ref{se:3.4} we assemble all pieces and present the result for the partition function.

\subsection{Second variation of the action}\label{se:3.1}

The second variation of the classical action around the (Euclidean) Lobachevsky background \eqref{eq:background} is given by
\begin{equation}
\delta^{\left(2\right)}S_{\textrm{\tiny CSG}}=\int\extd^{3}x\sqrt{\bar g}\,h^{\alpha\beta}\delta C_{\alpha\beta}\,,\label{eq:2.04}
\end{equation}
where we dropped the overall normalization constant in the action and $h^{\alpha\beta}$ is the metric fluctuation around the classical background. The second variation of the Cotton tensor is given by
\begin{eqnarray}
\delta C_{\alpha\beta} & = & -\frac{1}{2}\epsilon_{\,\,\,\,\,\,\alpha}^{\mu\nu}\nabla_{\mu}\left[\left(\nabla^{2}+R\right)h_{\nu\beta}+\left(\nabla_{\nu}\nabla_{\beta}+2R_{\nu\beta}\right)h_\ga^\ga-6\, h_{\gamma(\nu}R_{\,\,\beta)}^{\gamma}-2\nabla_{(\nu}\left(\nabla\cdot h\right)_{\beta)}\right]\nonumber \\
 &  & -\frac{1}{2}\epsilon_{\,\,\,\,\,\,\alpha}^{\mu\nu}R_{\mu}^{\ga}\left(\nabla_{\ga}h_{\nu\beta}-\nabla_{\nu}h_{\beta\ga}-\nabla_{\beta}h_{\nu\ga}\right)+\left(\alpha\leftrightarrow\beta\right)\,.\label{eq:2.05}
\end{eqnarray}

We decompose the field $h_{\mu\nu}$ into its transverse-traceless (TT), `trace' and gauge modes:
\begin{equation}
h_{\mu\nu}=h_{\mu\nu}^{TT}+ h\,g_{\mu\nu}+2\nabla_{(\mu}\xi_{\nu)}\,.\label{dec}
\end{equation}
The $h_{\mu\nu}^{TT}$ harmonics satisfy the standard conditions:
\begin{equation}
\nabla^{\mu}h_{\mu\nu}^{TT}=0\qquad g^{\alpha\beta}h_{\alpha\beta}^{TT}=0\label{eq:TT}
\end{equation}
They play the role of the gauge-fixed modes $h_{\rm gf}$ introduced above.

Let $k$ be a unit vector tangential to the $S^1$. Then the variation of the Cotton tensor evaluated on TT fluctuations simplifies to
\begin{eqnarray}
\delta C_{\alpha\beta}^{TT} & = & -\frac{1}{2}\epsilon_{\,\,\,\,\,\,\alpha}^{\mu\nu}\nabla_{\mu}\left[\left(\nabla^{2}+2\right)h_{\nu\beta}^{TT}-2k^{\gamma}k_{\nu}h_{\gamma\beta}^{TT}-3k^{\gamma}k_{\beta}h_{\gamma\nu}^{TT}\right]\nonumber \\
 &  & -\frac{1}{2}\epsilon_{\,\,\,\,\,\,\alpha}^{\mu\nu}k_{\mu}k^{\gamma}\left(\nabla_{\gamma}h_{\nu\beta}^{TT}-\nabla_{\beta}h_{\gamma\nu}^{TT}\right) 
+\left(\alpha\leftrightarrow\beta\right),\label{eq:21}\end{eqnarray}
where we used the background identities \eqref{eq:lob43}.
Due to the diffeomorphism and Weyl invariance, $h$ and $\xi_\mu$ do not contribute to the variation (\ref{eq:21}).

Further, we make a Kaluza--Klein decomposition of the TT harmonics,
\eq{
h_{\mu\nu}^{TT}=\left(\begin{array}{cc}
h_{\tau\tau} & h_{\tau a}\\
h_{a\tau} & h_{ab}\end{array}\right)\,,
}{eq:angelinajolie}
where $\tau$ is the Euclidean time direction along the $S^1$ and $a,b=1,2$.
This split yields
\begin{align}
{\delta}C_{\tau\tau}^{TT} & =  -\epsilon^{ab}\nabla_{a}\left(\nabla^{2}-1\right)h_{\tau b}\label{eq:22}\\
{\delta}C_{\tau a}^{TT} & =  -\frac{1}{2}\epsilon^{bc}\nabla_{b}\left(\nabla^{2}+2\right)h_{ca}+\frac{1}{2}\epsilon_{a}^{\,\,\, b}\nabla_{\tau}\left(\nabla^{2}-1\right)h_{\tau b}-\frac{1}{2}\epsilon_{a}^{\,\,\, b}\nabla_{b}\left(\nabla^{2}-3\right)h_{\tau\tau}\label{eq:23}\\
{\delta}C_{ab}^{TT} & =  \frac{1}{2}\epsilon_{a}^{\,\,\, c}\left[\nabla_{\tau}\left(\nabla^{2}+3\right)h_{cb}-\nabla_{c}\nabla^{2}h_{\tau b}-\nabla_{b}h_{\tau c}\right]+\left(a\leftrightarrow b\right)\,.\label{eq:24}
\end{align}
Here $\nabla^2:= \nabla^\mu\nabla_\mu$. Later we shall also use $\Delta:=\nabla^a\nabla_a$.

After a long but otherwise straightforward algebra one may resolve the TT conditions (\ref{eq:TT}) and
express $h^{TT}$ in terms of a scalar $s$ and a pseudoscalar $p$
\begin{equation}
 h_{\mu\nu}^{TT}=h_{\mu\nu}^{(S)}(s)+h_{\mu\nu}^{(P)}(p) \,,\label{eq:hsp}
\end{equation}
where
\begin{subequations}
\label{eq:hS}
\begin{eqnarray}
&&h_{\tau\tau}^{(S)}(s)=-\Delta (\Delta-2)s\\
&&h_{a\tau}^{(S)}(s)=\nabla_a \partial_\tau (\Delta-2)s\\
&&h_{ab}^{(S)}(s)=-\nabla_a\nabla_b (\Delta +2\partial_\tau^2) s + g_{ab} \Delta (\nabla^2-1)s
\end{eqnarray}
\end{subequations}
and
\begin{subequations}
\label{eq:hP}
\begin{eqnarray}
&&h_{\tau\tau}^{(P)}(p)=0\\
&&h_{a\tau}^{(P)}(p)=\epsilon_a^{\ b}\nabla_b  (\Delta-2)p\\
&&h_{ab}^{(P)}(p)=-\bigl(\epsilon_a^{\ c}\nabla_b\nabla_c +\epsilon_b^{\ c}\nabla_a\nabla_c\bigr)
\partial_\tau p \,.
\end{eqnarray}
\end{subequations}
Then one can demonstrate that
\begin{subequations}
 \label{eq:OPS}
\begin{eqnarray}
&&\delta C_{\mu\nu}^{TT}\big[h^{(P)}(p)\big]=h^{(S)}_{\mu\nu}\big( -(\nabla^2-2)p\big)\\
&&\delta C_{\mu\nu}^{TT}\big[h^{(S)}(s)\big]=h^{(P)}_{\mu\nu}\big( [(\nabla^2-1)^2 +\frac12\,(\Delta-2)]s\big)\,.
\end{eqnarray}
\end{subequations}

Now we can calculate the second factor in the last term of (\ref{eq:pint}), which yields the one-loop determinant of the gauge-fixed fluctuations $h_{\rm gf}$
\begin{eqnarray}
 Z_{\rm TT}&=& \int \mathcal{D}h_{\mu\nu}^{TT} e^{-S_2(h^{TT})}\nonumber\\
&=& \Big[\det \big(-(\nabla^2-2)\big)\big((\nabla^2-1)^2 +\frac12\,(\Delta-2)\big)\Big]^{-1/2}_0 \,,\label{TTint}
\end{eqnarray}
where the subscript $0$ means that the determinant is calculated for $\mathbb{H}^2$ scalars (rather than tensors or vectors).

\subsection{Gauge modes}\label{se:3.2}

The ghost factor is equal to the Jacobian appearing in the path integral measure after the change of variables (\ref{dec}),
\begin{equation}
 \mathcal{D}h_{\mu\nu}=Z_{\rm gh} \mathcal{D}h_{\mu\nu}^{TT} \mathcal{D} h\mathcal{D}\xi_\mu \,.\label{Zgh}
\end{equation}
To compute $Z_{\rm gh}$ it is convenient to Kaluza--Klein decompose $\xi_\mu$ and further decompose the vector part into exact and co-exact contributions,
\begin{eqnarray}
&&\xi_\mu =\xi^{(1)}_\mu + \xi^{(2)}_\mu +\xi^{(3)}_\mu\\
&&\xi^{(1)}_\tau=u\qquad \xi^{(1)}_a=0\\
&&\xi^{(2)}_\tau=0\qquad \xi^{(2)}_a=\partial_a v\\
&&\xi^{(3)}_\tau=0\qquad \xi^{(3)}_a=\epsilon_a^{\ b}\partial_b w
\end{eqnarray}
with three scalars $u,v,w$. 
Each change of variables generates a Jacobian factor
\begin{eqnarray}
&&\mathcal{D}h_{\mu\nu}=J_1 \mathcal{D}h_{\mu\nu}^{TT} \mathcal{D} h \mathcal{D}u \mathcal{D} v
\mathcal{D}\xi^{(3)}\nonumber \\
&&\mathcal{D}\xi_\mu =J_2 \mathcal{D}u \mathcal{D} v \mathcal{D} \xi^{(3)}\,,\label{J1J2}
\end{eqnarray}
so that the ghost contribution to the one-loop partition function is the ratio of these Jacobians.
\begin{equation}
 Z_{\rm gh}=J_1/J_2 \label{ZJJ}
\end{equation}
Each of these factors can be calculated by using the normalization condition for the path integral
measure.
Then,
\begin{eqnarray}
1&=&\int \mathcal{D}h_{\mu\nu} \exp(-\langle h,h \rangle ) \nonumber\\
&=&\int J_1 \mathcal{D}h_{\mu\nu}^{TT} \mathcal{D} h \mathcal{D}u \mathcal{D} v
\mathcal{D}\xi^{(3)} \exp \Big( -\int\extd^3x\sqrt{g}\, h_{\mu\nu}h^{\mu\nu} \Big)\nonumber\\
&=&\int J_1 \mathcal{D} h \mathcal{D}u \mathcal{D} v
\mathcal{D}\xi^{(3)} \exp \Big( -\int\extd^3x\sqrt{g} \,(h,u,v,\xi^{(3)}) A (h,u,v,\xi^{(3)})^t \Big)
\nonumber
\end{eqnarray}
where $t$ means a transposition, and
\begin{equation}
 A=\left( \begin{array}{cccc}
           3 & 2\partial_\tau & 2\Delta & 0 \\
-2\partial_\tau & -4\partial_\tau^2 -2 \Delta & -2\Delta\partial_\tau & 0 \\
2\Delta & 2 \Delta \partial_\tau & 2\Delta (\partial_\tau^2 + 2\Delta -2) & 0 \\
0 & 0 & 0 & -2(\nabla^2-1) 
          \end{array} \right) \,.
\end{equation}
Therefore, the Jacobian factor $J_1$ yields
\begin{eqnarray}
&& J_1=[\det A]^{1/2}\label{J1}\\
&&\quad =\Big[\det\big(-\Delta\big)_0 \det\big((\nabla^2-1)^2 +(1/2)(\Delta-2)\big)_0
\det\big(-(\nabla^2-1)\big)^T_1\Big]^{1/2}\,.\nonumber
\end{eqnarray}
The subscript $0$ ($1$) means that the determinant is calculated on $\mathbb{H}^2$-scalars
(vectors). The superscript $T$ means that the vectors are transverse. 
By using the identity
\begin{equation}
 (\nabla^2-1)\epsilon_a^{\ b} \nabla_b w=\epsilon_a^{\ b}\nabla_b (\nabla^2 -2) w
\end{equation}
one can rewrite the vector determinant in (\ref{J1}) as a scalar determinant,
\begin{equation}
 \det\big(-(\nabla^2-1)\big)^T_1=\det\big(-(\nabla^2-2)\big)_0 \label{10}
\end{equation}

The Jacobian factor $J_2$ can be calculated similarly. The identity
\begin{eqnarray}
1&=&\int \mathcal{D}\xi_\mu \exp\Big( -\int\extd^3x\, \xi_\mu \xi^\mu \Big)\nonumber\\
&=&\int J_2 \mathcal{D}u \mathcal{D} v \mathcal{D} \xi^{(3)} \exp\Big( -\int\extd^3x\sqrt{g} \,
(u^2 + v (-\Delta)v + \xi^{(3)}_\mu \xi^{(3)\mu}) \Big)
\end{eqnarray}
yields
\begin{equation}
 J_2=[\det (-\Delta)]_0^{1/2} \,.\label{J2}
\end{equation}
Therefore, the one-loop ghost determinant simplifies to
\begin{equation}
 Z_{\rm gh}=\Big[\det\big(-(\nabla^2-2)\big)_0 \det\big((\nabla^2-1)^2 +\frac12\,(\Delta-2)\big)_0 \Big]^{1/2}\,.\label{eq:Zgh}
\end{equation}
The ghost determinant \eqref{eq:Zgh} is formally the inverse of the physical determinant \eqref{TTint}, which appears to suggest a trivial 1-loop partition function. However, as we show below it is crucial to take into account the different boundary behavior of physical and ghost modes, as a consequence of which the 1-loop partition function becomes non-trivial.

\subsection{Boundary conditions}\label{se:3.3}

Let us define the boundary conditions on the scalar fields $s,p,h,u,v,w$ consistent with the Lobachevsky boundary conditions \eqref{eq:bcs}, \eqref{eq:lob5} on $h_{\mu\nu}$. In this analysis one can use the asymptotic version of the metric \eqref{eq:background}
\eq{
\extd\bar s^2 = \frac{\extd y^2}{y^2}+\frac{\extd\varphi^2}{y^2} + \extd \tau^2
}{eq:lob20}
with the Christoffel connection $\bar\Gamma^\varphi{}_{\varphi y}=-\bar\Gamma^y{}_{\varphi\varphi}=\bar\Gamma^y{}_{yy}=-y^{-1}$.
The corresponding two-dimensional Laplace operator is just
$\bar\Delta=y^{2}\left(\partial_{\varphi}^{2}+\partial_{y}^{2}\right)$.

After long but straightforward calculations we obtain the boundary conditions on the scalar fields $s$ and $p$,
\begin{equation}
 s=s_{-1}(\tau,\varphi)y^{-1}+s_0(\tau,\,\varphi) + \mathcal{O}(y)\qquad 
 p=p_{-1}(\tau,\varphi)y^{-1}+p_0(\tau,\,\varphi) + \mathcal{O}(y)\,, \label{bcsp}
\end{equation}
where $\mathcal{O}(y)$ means \emph{any} power (possibly non-integer) of $y$ that is equal or greater 
than one.\footnote{At first glance the Lobachevsky boundary conditions also seem to allow modes of the form $s\sim y \ln(y) \sigma(\varphi)$. 
However, imposing the asymptotic equation of motion \eqref{eq:app1} for these fluctuations requires vanishing $\sigma$. Thus, we impose the boundary conditions \eqref{bcsp} with no loss of essentiality.}
The leading contributions $s_{-1}$, $p_{-1}$, $s_0$ and $p_0$ are asymptotically growing and
asymptotically constant modes. 
The growing and constant terms are isolated solutions that will play an important role below.
In appendix \ref{app:sp} we discuss which physical states are generated by these modes.

Let us now turn to the gauge sector. One can easily find the boundary conditions for gauge modes
in the decomposition (\ref{dec}),
\begin{equation}
 h=\mathcal{O}(y)\qquad \xi_\tau=\mathcal{O}(y)\qquad \xi_\varphi =\mathcal{O}(1)
\qquad \xi_y =\mathcal{O}(1)\,.\label{bchxi}
\end{equation}
Thus, all the gauge scalars are of the same order,
\begin{equation}
 h,\, u,\, v,\, w =\mathcal{O}(y)\,.\label{bchuvw}
\end{equation}
The scalar $h$ ($u$) [$v$] \{$w$\} corresponds to a multiple of $\omega_1$ ($T_1$) [$H_1$] \{$L_1$\} in the notation of section \ref{se:2.3}, and thus manifestly generates small gauge transformations.
Isolated asymptotically constant solutions are allowed for $v$ and $w$, but they do not generate
independent solutions for $\xi_a$ and have to be discarded. To see this, let us take $v$, $w$ in the form of a
Taylor series
  \eq{
v=\sum_{i=0}^{n}v_{i}\left(\varphi\right)y^{i}\qquad w=\sum_{i=0}^{n}w_{i}\left(\varphi\right)y^{i}\,.
}{eq:newlabel1}
Then,
\begin{equation}
\xi_{\varphi}  =  
\partial_{\varphi}v-\partial_{y}w= \left[v_{0}'-w_{1}\right]+\sum_{i=1}^{n}\left[v_{i}'-\left(i+1\right)w_{i+1}\right]y^{i}\,.
\end{equation}
and
\begin{equation}
\xi_{y} =  \partial_{y}v+\partial_{\varphi}w=
\left[v_{1}+w_{0}'\right]+\sum_{i=1}^{n}\left[\left(i+1\right)v_{i+1}+w_{i}'\right]y^{i}\,.
\end{equation}
From these expressions it is clear that one can obtain arbitrary Taylor expansions for $\xi_\varphi$ and
$\xi_y$ by adjusting the Taylor coefficients of $v$ and $w$ with the constraints $v_0=w_0=0$.

\newcommand{\whatever}{f}

Let us analyze the isolated modes $s_{-1}$, $p_{-1}$, $s_0$ and $p_0$. To this end, it is convenient to relax
for a while the Lobachevsky boundary conditions \eqref{eq:bcs} and extend the space of metric perturbations to all square
integrable TT modes. Such modes are generated through the relations (\ref{eq:hP}) and
(\ref{eq:hS}) by the scalar modes $\bar s$ and $\bar p$ satisfying the boundary conditions
$\bar s=s_{-1}y^{-1}+s_0+ o(y^{1/2})$ and $\bar p=p_{-1}y^{-1}+ p_0+ o(y^{1/2})$. In other words, 
square integrable TT
modes are generated by square integrable scalar modes and by isolated asymptotically growing  
and constant modes $s_{-1}$, $p_{-1}$,
$s_0$ and $p_0$. This implies that the TT fields generated by the isolated modes are orthogonal to the TT fields
generated by square integrable scalars. 

Let us consider the asymptotically constant modes.
Since $L^2(\mathbb{H}^2\times S^1)$ is a closure of the space of smooth rapidly 
decaying functions $\mathcal{S}(\mathbb{H}^2\times S^1)$, one has the conditions
\begin{eqnarray}
&& 0=\int\extd^3x \sqrt{g} \,h_{\mu\nu}^{(S)}(s_0)h^{(S)\mu\nu}(\tilde s)\label{orts}\\
&& 0=\int\extd^3x \sqrt{g} \,h_{\mu\nu}^{(P)}(p_0)h^{(P)\mu\nu}(\tilde p)\label{ortp}
\end{eqnarray}
for $\tilde p,\tilde s\in \mathcal{S}(\mathbb{H}^2\times S^1)$. Using the Schwartz space has an obvious advantage that
one can integrate by parts in (\ref{orts}) and (\ref{ortp}) thus arriving at
\begin{eqnarray}
&& 0=\int\extd^3x \sqrt{g}\, \tilde s \big(2\nabla^2(\nabla^2-2)+\Delta\big)\Delta (\Delta -2) s_0\\\label{orts2}
&& 0=\int\extd^3x \sqrt{g} \,\tilde p\, \Delta (\Delta -2) (\nabla^2-2) p_0\label{ortp2}\,.
\end{eqnarray}
By these equations, $s_0$ and $p_0$ have to satisfy the differential equations
\begin{eqnarray}
&&0=\big(2\nabla^2(\nabla^2-2)+\Delta\big)\Delta (\Delta -2) s_0  \label{orts3}\\
&&0=\Delta (\Delta -2) (\nabla^2-2) p_0 \label{ortp3}
\end{eqnarray}
and behave as a constant at the boundary. One can show that for any given dependence on
$\varphi$ and $\tau$ the problems above may have at most one solution (up to an overall constant).
Indeed, suppose that there are two modes, $p_0^{(1)}$ and $p_0^{(2)}$ satisfying (\ref{ortp2})
such that $p^{(1,2)}_0=P(\varphi,\tau)+\mathcal{O}(y)$. Then the difference 
$p_0^{(1)}-p_0^{(2)}=\mathcal{O}(y)$ and also satisfies (\ref{ortp2}). On the $\mathcal{O}(y)$ fields
the operators on the right hand side of (\ref{ortp2}) are positive and invertible \cite{Camporesi:1994ga}.
Consequently, $p_0^{(1)}-p_0^{(2)}=0$.
Therefore, one can simply try zero modes of the operators in (\ref{orts3}) and (\ref{ortp3})
until this solution is found. The solutions are identical for $s_0$ and $p_0$ and read
\begin{equation}
 s_0,\,p_0=\left[ \frac {\sinh (\rho)}{1+\cosh (\rho) }\right]^{|h|} e^{-ih\varphi} \,\whatever^{s,p}_0(\tau) \,,\label{p0s0}
\end{equation}
where $\whatever^{s,p}_0$ are arbitrary functions of $\tau$ and $h$ is an integer. These solutions satisfy
\begin{equation}
 \Delta s_0 = \Delta p_0=0 \label{D00}
\end{equation}
and obey a regularity condition at the origin, $\lim_{\rho\to 0} |s_0|,\, |p_0| < \infty$.
One can easily check that the corresponding TT modes are non-zero except for $h=0$. 

By repeating the same analysis for $s_{-1}$ and $p_{-1}$ we arrive at
\begin{equation}
 s_{-1},\,p_{-1}=\left[ \frac {\sinh (\rho)}{1+\cosh (\rho) }\right]^{|h|} \big(|h| + \cosh (\rho)\big)
 e^{-ih\varphi} \,\whatever^{s,p}_{-1}(\tau) \,,\label{pm1sm1}
\end{equation}
and
\begin{equation}
(\Delta -2)s_{-1}=(\Delta -2)p_{-1}=0\,.\label{Dm2}
\end{equation}
Non-zero TT modes are generated for $|h|\ge 2$.

The modes (\ref{p0s0}), (\ref{pm1sm1})
have remarkable properties
\begin{equation}
h^{(P)}_{\mu\nu}(i\partial_\tau s_0)=h^{(S)}_{\mu\nu}(s_0)\,,\qquad
h^{(S)}_{\mu\nu}(\partial_\tau s_{-1})=h^{(P)}_{\mu\nu}(i(\partial_\tau^2+1)s_{-1})
\,.\label{rem}
\end{equation}
Since the function $f^{s,p}_0(\tau)$ in (\ref{p0s0}) is arbitrary, this implies that the $s_0$ and $p_0$
modes generate the same metric fluctuations. To avoid double counting in the path integral, we should
keep one set of the modes only. The same applies to $s_{-1}$ and $p_{-1}$.

The calculation above also demonstrates that the tensor modes generated by $p_{-1}$ or $s_{-1}$ 
and by $p_0$ or $s_0$
cannot be obtained from $\mathcal{O}(y)$ scalars (as happened with the $v$ and $w$ gauge modes)
since this would contradict the orthogonality conditions (\ref{orts}) and (\ref{ortp}) and similar conditions
for $s_{-1}$ and $p_{-1}$. 

\subsection{Aspects of the Lobachevsky $\leftrightarrow$ field theory map}

It is instructive to perform an analysis similar to section 4 of \cite{Maldacena:1998bw} where the correspondence between states in AdS and the conformal field theory was studied. As a first step, we write the 3-dimensional d'Alembert operator on scalar fields as sum of quadratic Casimirs, using the explicit form of the Killing vectors \eqref{eq:referee1} of the Lobachevsky background.
\eq{
\nabla^2 = T_0^2 + L_0^2 - \frac12\, \big(L_{+1} L_{-1} + L_{-1} L_{+1}\big) = - \partial_t^2 + \partial_\rho^2+\coth\rho\,\partial_\rho + \frac{1}{\sinh^2\!\rho}\,\partial_\varphi^2
}{eq:referee3}
Similarly, the 2-dimensional Laplacian on scalar fields is just the quadratic Casimir of the $SL(2)$ part of the isometry algebra.
\eq{
\De = L_0^2 - \frac12\, \big(L_{+1} L_{-1} + L_{-1} L_{+1}\big) = \partial_\rho^2+\coth\rho\,\partial_\rho + \frac{1}{\sinh^2\!\rho}\,\partial_\varphi^2
}{eq:referee4}
This means that the isometry algebra can be used to classify solutions of the wave or Laplace equation, like the ones we have encountered above.
We can then label states $|\psi\rangle$ in the field theory according to their $U(1)$  and $SL(2)$ weights $(j,\,h)$.
\eq{
T_0|\psi\rangle = j|\psi\rangle\qquad L_0|\psi\rangle = h|\psi\rangle
}{eq:referee5}
In what follows, the $t$- (or Euclidean $\tau$-) dependence will not play any significant role, which is why we disregard the weights $j$.

Now suppose that $|\psi\rangle$ is a primary state in the sense that $L_1|\psi\rangle=0$. Using the separation Ansatz $|\psi\rangle=f(\tau)\,e^{-ih\varphi}\,F(\rho)$ we find that the function $F$ satisfies
\eq{
F(\rho) = \frac{F_0}{(\sinh\rho)^h}
}{eq:referee6}
where $F_0$ is some normalization constant.
Finiteness of the primary at small $\rho$ requires $h \leq 0$.
Compatibility with our boundary conditions \eqref{bcsp}, which we call ``normalizability'', leads to the inequality $h \geq -1$.
In conclusion, finite, normalizable primaries with integer weights must have either weight $h=0$ or weight $h=-1$. This explains from a field theory point of view why we have found exactly two towers of (perturbative) states on the gravity side, \eqref{p0s0} and \eqref{pm1sm1}.

If a primary state represents a scalar field of mass $m$, $(\De-m^2)|\psi\rangle=0$, then we obtain from the identity $\De=L_0(L_0-1)-L_{-1}L_1$ a relation between the mass $m$ and the allowed weights of the primary:
\eq{
h = \frac12\,\big(1\pm\sqrt{1+4m^2}\big)
}{eq:referee7}
Note that $h$ is real as long as the 2-dimensional Breitenlohner--Freedman bound is satisfied, $m^2\geq m^2_{\textrm{\tiny BF}} = -\tfrac14$.
Imposing finiteness at small $\rho$ picks the lower sign in equation \eqref{eq:referee7} and requires non-negative $m^2$. Normalizability \eqref{bcsp} leads to the inequality $m^2\leq 2$.
Thus, finite, normalizable scalar fields must have a mass in the range $0\leq m^2\leq 2$, concurrent with \eqref{D00} and \eqref{Dm2}, which saturate the respective bounds.

The analysis above allows us to discuss the algebraic properties of the modes \eqref{p0s0} and \eqref{pm1sm1}.
The modes \eqref{p0s0} [the modes \eqref{pm1sm1}] obey the primary condition \eqref{eq:referee6} only for vanishing weight, $h=0$ [weight $h=-1$]. This is consistent with the results we just derived.
Denoting these modes as $s_{0,-1},\,p_{0,-1}=|0/-1,\,h\rangle^{s,p}$ we obtain the following algebraic relations.
\begin{align}
 L_{-1} |0,\,h\rangle^{s,p} &= h\,|0,\,h+1\rangle^{s,p} &  L_{-1} |-1,\,h\rangle^{s,p} &= (h-1)\,|-1,\,h+1\rangle^{s,p} \\
 L_1 |0,\,h\rangle^{s,p} &= h\,|0,\,h-1\rangle^{s,p} & L_1 |-1,\,h\rangle^{s,p} &= (h+1)\,|-1,\,h-1\rangle^{s,p}  
\end{align}
Thus, acting with the raising (lowering) operator $L_{-1}$ ($L_1$) on a state of weight $h$ leads in general to a state of weight $h+1$ ($h-1$), as expected.

\subsection{Partition function}\label{se:3.4}

We have now all the pieces available to assemble the result for the one-loop partition function \eqref{eq:pint} of conformal Chern--Simons gravity with Lobachevsky boundary conditions.

The contributions of TT modes and ghosts to the partition function, see (\ref{TTint}) and (\ref{eq:Zgh}),
are given by determinants of the same scalar operators, but the boundary conditions are different.
Therefore, there are non-compensated contributions of the boundary modes $p_0$ and $p_{-1}$ 
(or $s_0$ and $s_{-1}$).
To evaluate the contribution from $p_0$, we compute
\begin{equation}
\delta C_{\mu\nu}^{TT} \big[ h^{(P)}(p_0)\big]=h_{\mu\nu}^{(S)} \big( -(\partial_\tau^2-2)p_0\big)
=h_{\mu\nu}^{(P)}\big( -i\partial_\tau (\partial_\tau^2-2) p_0 \big)\,,
\end{equation}
where we used (\ref{eq:OPS}), (\ref{D00}) and (\ref{rem}). This yields
\begin{equation}
 Z_0=\big[\det (-i\partial_\tau)(\partial_\tau^2-2) \big]^{-1/2}_{p_0}\,.\label{eq:illdet}
\end{equation}
The operator in (\ref{eq:illdet}) is just a square root of the operator appearing in (\ref{TTint}) for the
harmonic scalars, as may be anticipated. Similarly, for the $p_{-1}$ mode we have
\begin{equation}
\delta C_{\mu\nu}^{TT}\big[ h^{(P)}(p_{-1})\big]=h^{(S)}(-\partial_\tau^2p_{-1})=
h^{(P)}(-i\partial_\tau (\partial_\tau^2+1)p_{-1})
\end{equation}
yielding
\begin{equation}
Z_{-1}=\big[ \det(-i\partial_\tau)(\partial_\tau^2+1) \big]^{-1/2}_{p_{-1}}\,.\label{eq:illm1}
\end{equation}
The full 1-loop partition function is
\begin{equation}
Z=Z_{0}Z_{-1}\,.\label{complete}
\end{equation}
The complete and explicit cancellation of all bulk modes is a remarkable property of conformal Chern--Simons gravity with Lobachevsky boundary conditions. We have thus achieved an explicit separation between bulk modes and boundary modes.

However, there is an infinite degeneracy in the $SL(2)$ weight $h$ [see \eqref{p0s0}, \eqref{pm1sm1}], so that it is not clear how the determinants \eqref{eq:illdet} and \eqref{eq:illm1} can be defined. We shall comment in the concluding section \ref{se:5} on a possible resolution of this problem.

\section{Discussion}\label{se:5}

In this paper we made the first steps to study Lobachevsky holography. We proposed Lobachevsky boundary conditions \eqref{eq:bcs} and implemented them successfully in conformal Chern--Simons gravity \eqref{eq:gCS10}. We constructed for this theory the canonical boundary charges and proved that they are non-trivial, integrable, finite and conserved. We calculated these charges for non-perturbative states in section \ref{app:np}. The asymptotic symmetry algebra \eqref{eq:lob7} contained an affine $\hat u(1)$ and a Virasoro algebra with positive central charge \eqref{eq:lob18}. We then focused on the one-loop partition function and calculated it. After several technical steps, including the careful consideration of boundary conditions, we managed to obtain a clear separation between bulk and boundary modes in the final result \eqref{complete}. However, we were not able to evaluate the determinants appearing in \eqref{eq:illdet} and \eqref{eq:illm1} due to an infinite degeneracy coming from the solutions for the boundary modes \eqref{p0s0}, \eqref{pm1sm1} which are labeled by an integer $h$. We left this issue as an open problem and address now its possible resolution.

The degeneracy probably can be removed by considering higher order terms in the action beyond the quadratic level. If true, this would imply that the partition function is not one-loop exact. The relevance of fluctuations that do not solve the linearized equations of motion actually is expected from the result for the conserved boundary charges \eqref{eq:lob6}, \eqref{eq:lob7}, which also depend on fluctuations that do not solve the linearized equations of motion. Given that the charges depend on the linearized modes quadratically, one may expect that the cubic action resolves this issue already.

Despite of the technical difficulties encountered at 1-loop, there is useful information we can glean from the modes that contribute to the physical part of the partition function \eqref{complete}. The modes $s_0,\, p_0$ \eqref{p0s0} should correspond to the descendants of the vacuum generated by the $\hat u(1)$-current algebra generators $T_{-n}$, with positive integer $n$, since they generate non-vanishing $T$-charges [see \eqref{app:B1} plugged into \eqref{eq:diffcharge}] and are zero (non-zero) for $n=0$ ($n\neq 0$). Similarly, the modes  $s_{-1},\, p_{-1}$ \eqref{pm1sm1} should correspond to the Virasoro descendants of the vacuum generated by $L_{-n-1}$, with positive integer $n$, since they generate vanishing $T$-charges, non-vanishing $L$-charges (though the evaluation of the latter is not meaningful at linearized level) and are zero (non-zero) for $n=-1,\, 0,\, 1$ ($|n|\geq 2$). We have also exhibited some additional aspects of the Lobachevsky/field theory correspondence. In particular, we have shown that the asymptotic modes transform properly under the isometries of Lobachevsky space and that imposing normalizability and finiteness leads exactly to the two towers of perturbative states that we found on the gravity side in our 1-loop calculation.

It is also of interest to understand the field theoretic interpretation of the additional non-perturbative states in section \ref{app:np} and of the absence of black hole solutions, whose presence is usually required for modular invariance of the partition function \cite{Maloney:2007ud}.

We mention finally that there is a plethora of other topological theories where Lobachevsky holography can be implemented, namely any three-dimensional spin-$n$ theory with some non-principal embedding of $sl(2)$ \cite{Gary:2012ms}. It is conceivable that the problematic issues we encountered above in the calculation of the one-loop partition function are absent for some of (or even all) these theories, since at least for the simplest spin-3 example the canonical charges turn out to be linear in the fields (up to a Sugawara-term) \cite{Afshar:2012nk}. 

\acknowledgments

We thank Hamid Afshar, Branislav Cvetkovic, St{\'e}phane Detournay, Michael Gary, Radoslav Rashkov, Max Riegler and Simon Ross for discussions.

MB was supported by FAPESP.
SE, DG and NJ were supported by the START project Y435-N16 of the Austrian Science Fund (FWF) and by the FWF project P21927-N16.
HG acknowledges the support from Universit\'a degli Studi di Torino, and
INFN, Sezione di Torino. DV was supported by CNPq and FAPESP.
He also acknowledges travel support from the FWF project Y435-N16.

\appendix

\section{Asymptotic equations of motion} \label{app:eom}

Only three of the asymptotic equations of motion are needed to prove conservation of the charges.\footnote{%
In the twelve leading and subleading equations of motion there is only one linearly independent non-trivial equation in addition to \eqref{eq:app1}-\eqref{eq:app3}, namely the subleading equation $\textrm{eom}_{yy}=0$, which we do not present here since it is not needed to prove the conservation of the charges. 
} They are given by $\textrm{eom}_{\mu\nu}=0$, with
\begin{align}
\textrm{eom}_{t\varphi} =&\;\frac{1}{2} \gtt
+\frac{1}{4} \gff + \frac{1}{4} \gyy -\partial_t \gty  + \frac{1}{4} \partial_t^2 \gyy - \frac{1}{4} \partial_t^2\gff \label{eq:app1} \,,\\
\textrm{eom}_{ty} =&\; \frac{\partial}{\partial t}\gtf - \frac{1}{2}\frac{\partial^2}{\partial t^2}\gfy \,,\label{eq:app2} \\
\textrm{eom}_{\varphi y} =&\; \big(1-\tfrac14\,\partial_t^2\big)\,\partial_t \partial_\varphi \gfy \nonumber \\
& -\tfrac32\,\partial_t\htt+\tfrac32\,\partial_t\hty-\big(1-\tfrac14\,\partial_t^2\big)\,\partial_t\hff-\big(\tfrac12+\tfrac14\,\partial_t^2\big)\,\partial_t\hyy \nonumber \\
& -\gty \big[\big(1-\partial_t^2\big)\,\partial_t\gty+\big(1-\tfrac14\,\partial_t^2\big)\,\partial_t^2\gff+\big(\tfrac12+\tfrac14\,\partial_t^2\big)\,\partial_t^2\gyy\big] \nonumber \\
& -\tfrac14\,\big[\big(1-\partial_t^2\big)\,\partial_t\gty\big]\,\big(\partial_t\gff-\partial_t\gyy\big) + \tfrac34\,\partial_t^2\gty\,\big(\gff+\gyy\big) \nonumber \\
& + \tfrac{1}{16}\,\big[\gff\,\big(9\partial_t\gff-5\partial_t\gyy-4\partial_t^3\gyy\big) \nonumber \\ 
& \qquad\; -\gyy\,\big(5\partial_t\gff-8\partial_t^3\gff-\partial_t\gyy+4\partial_t^3\gyy\big) \nonumber \\
& \qquad\; -\big(\partial_t\gff-\partial_t\gyy\big)\,\big(10\partial_t^2\gff-\partial_t^4\gff+2\partial_t^2\gyy+\partial_t^4\gyy\big)\big] \nonumber \\
& + \gfy \,\big(1-\tfrac14\,\partial_t^2\big)\,\partial_t\gfy\,.
\label{eq:app3}
\end{align}
We note that the last equation is actually a subleading equation in the large $y$ expansion, so this equation does not arise at the linearized level. Moreover, it relates terms linear in $\gfy$ with second order terms linear in $g^{(2)}$ or bi-linear in $g^{(1)}$, with a structure similar to the expression for the Virasoro boundary charges \eqref{eq:lob7}. 

\section{Physical states generated by scalar modes} \label{app:sp}

Let us switch on the modes generated by $s=s_{-1}(\tau,\, \varphi)/y + s_0(\tau,\, \varphi) + s_1(\tau,\, \varphi)\,y + s_2(\tau,\,\varphi)\,y^2+ o(y^2)$ and calculate explicitly the metric fluctuations \eqref{eq:hS}. In the conventions of sections \ref{se:2}, \ref{se:3} we obtain
\begin{align}
 g_{\tau\tau}^{(1)} &= 0 & g_{\tau\varphi}^{(1)} &= -2\partial_\tau\partial_\varphi s_0 & g_{\tau y}^{(1)} &= -2\partial_\tau\tilde s \label{app:B1} \\
 g_{\varphi\varphi}^{(1)} &= 2\partial_\tau^2\tilde s & g_{\varphi y}^{(1)} &= -2\partial_\tau^2\partial_\varphi s_0 & g_{yy}^{(1)} &= -2\partial_\tau^2\tilde s 
\label{app:B2}
\end{align}
where $\tilde s=s_1-\tfrac12\,(\partial_\varphi^2+\tfrac12)\,s_{-1}$, and
\begin{align}
 g_{\tau\tau}^{(2)} &= 0 & g_{\tau\varphi}^{(2)} &= -2\partial_\tau\partial_\varphi\tilde s & g_{\tau y}^{(2)} &= 2\partial_\tau\partial_\varphi^2 s_0 \\
 g_{\varphi\varphi}^{(2)} &=  \check s & g_{\varphi y}^{(2)} &= -4\partial_\varphi\hat s & g_{yy}^{(1)} &= - \check s
\end{align}
where $\hat s=\partial_\tau^2 s_1+\tfrac12\,(\partial_\varphi^2+\tfrac12\,\partial_\tau^2+1)s_{-1}$ and $\check s=\partial_\varphi^2(3-\partial_\tau^2) s_0+6(\partial_\tau^2+1)s_2$.
Similarly, modes generated by $p=p_{-1}(\tau,\,\varphi)/y + p_0(\tau,\, \varphi) + p_1(\tau,\, \varphi)\,y + p_2(\tau,\,\varphi)\,y^2 + o(y^2)$ yield
\begin{align}
 g_{\tau\tau}^{(1)} &= 0 & g_{\tau\varphi}^{(1)} &= 2\tilde p & g_{\tau y}^{(1)} &= -2\partial_\varphi p_0 \\
 g_{\varphi\varphi}^{(1)} &= 2\partial_\tau\partial_\varphi p_0 & g_{\varphi y}^{(1)} &= 2\partial_\tau\tilde p & g_{yy}^{(1)} &= -2\partial_\tau\partial_\varphi p_0 
\end{align}
where $\tilde p=p_1-\tfrac12\,(\partial_\varphi^2+\tfrac12) p_{-1}$, and
\begin{align}
 g_{\tau\tau}^{(2)} &= 0 & g_{\tau\varphi}^{(2)} &= -2\partial_\varphi^2 p_0 & g_{\tau y}^{(2)} &=  -2\partial_\varphi\tilde p\\
 g_{\varphi\varphi}^{(2)} &= 4\partial_\tau\partial_\varphi \hat p & g_{\varphi y}^{(2)} &= -\partial_\tau\check p & g_{yy}^{(2)} &= -4\partial_\tau\partial_\varphi \hat p
\end{align}
where $\hat p=p_1+\tfrac14\, p_{-1}$ and $\check p=\partial_\varphi^2 p_0 - 6p_2$.
The function $\tilde p$ (the function $\partial_\tau\tilde s$) generates the same leading metric fluctuations $g^{(1)}_{\mu\nu}$ as the function $-\partial_\tau\partial_\varphi s_0$ (the function $\partial_\varphi p_0$).

To discuss physical states at the linearized level we set to zero $\tilde s$ and $\tilde p$, and switch on alternately $s_0$ or $p_0$. We start with states generated solely by $s_0$.
According to \eqref{eq:trans} such states are not physical but pure gauge at linearized level if the condition $(\partial_\tau^2-2)\partial_\tau\partial_\varphi s_0=0$ is met. The on-shell condition \eqref{eq:app2} requires $(\partial_\tau^2-2)\partial_\tau^2\partial_\varphi s_0=0$. Therefore, physical states with non-vanishing $T$-charges generated by $s_0$ at linearized level are $\varphi$-dependent functions that solve
$\partial_\tau^2 s_0 = 0$ with $\partial_\tau s_0 \neq 0$.
Now we consider states generated solely by $p_0$.
Pure gauge modes at the linearized level have to obey the condition $(\partial_\tau^2 - 2)\partial_\tau\partial_\varphi p_0 = 0$. The on-shell condition \eqref{eq:app1} requires $\partial_\tau\partial_\varphi p_0 = 0$. Therefore, no physical states are generated by $p_0$ at linearized level and their $T$-charges vanish.


\providecommand{\href}[2]{#2}\begingroup\raggedright\endgroup

\end{document}